\pdfoutput=1
\documentclass[11pt]{article}

\usepackage[final]{acl}

\usepackage{times}
\usepackage{latexsym}
\usepackage{amsmath} 
\usepackage[T1]{fontenc}
\usepackage[utf8]{inputenc}

\usepackage{microtype}

\usepackage{inconsolata}

\usepackage{graphicx}

\title{Design Proteins Using Large Language Models: Enhancements and Comparative Analyses}

\author{
 \textbf{Kamyar Zeinalipour\textsuperscript{1}},
 \textbf{Neda Jamshidi\textsuperscript{1}},
 \textbf{Monica Bianchini\textsuperscript{1}},
 \textbf{Marco Maggini\textsuperscript{1}},
\\
 \textbf{Marco Gori\textsuperscript{1}},
\\
\\
  \textsuperscript{1}University of Siena, DIISM, Via Roma 56, 53100 Siena, Italy
\\
  \small{
   \textbf{Correspondence:} \href{kamyar.zeinalipour2@unisi.it}{kamyar.zeinalipour2@unisi.it}
  }
}

\begin{document}
\maketitle
\begin{abstract}
%In this paper, we introduce a family of large language models (LLMs), namely \texttt{P-Mistral}, \texttt{P-Llama2}, \texttt{P-Llama3}, and \texttt{P-gemma}, for protein generation. 
Pre-trained LLMs have demonstrated substantial capabilities across a range of conventional natural language processing (NLP) tasks, such as summarization and entity recognition. 
In this paper, we explore the application of LLMs in the generation of high-quality protein sequences. Specifically, we adopt a suite of pre-trained LLMs, including \texttt{Mistral-7B}\footnote{\href{https://huggingface.co/Kamyar-zeinalipour/P-Mistral-7B}{huggingface.co/Kamyar-zeinalipour/P-Mistral-7B}}, \texttt{Llama-2-7B}\footnote{\href{https://huggingface.co/Kamyar-zeinalipour/P-Llama2-7B}{huggingface.co/Kamyar-zeinalipour/P-Llama2-7B}}, \texttt{Llama-3-8B}\footnote{\href{https://huggingface.co/Kamyar-zeinalipour/P-Llama3-8B}{https://huggingface.co/Kamyar-zeinalipour/P-Llama3-8B}}, and \texttt{gemma-7B}\footnote{\href{https://huggingface.co/Kamyar-zeinalipour/P-gemma-7B/}{huggingface.co/Kamyar-zeinalipour/P-gemma-7B}}, to produce valid protein sequences. All of these models are publicly available.\footnote{\href{https://github.com/KamyarZeinalipour/protein-design-LLMs}{github.com/KamyarZeinalipour/protein-design-LLMs}} Unlike previous work in this field, our approach utilizes a relatively small dataset comprising 42,000 distinct human protein sequences. We retrain these models to process protein-related data, ensuring the generation of biologically feasible protein structures. Our findings demonstrate that even with limited data, the adapted models exhibit efficiency comparable to established protein-focused models such as ProGen varieties, ProtGPT2, and ProLLaMA, which were trained on millions of protein sequences. To validate and quantify the performance of our models, we conduct comparative analyses employing standard metrics such as pLDDT, RMSD, TM-score, and REU. Furthermore, we commit to making the trained versions of all four models publicly available, fostering greater transparency and collaboration in the field of computational biology.
\end{abstract}

\section{Introduction}

In recent years, the field of natural language processing (NLP) has achieved remarkable progress, particularly through the development and utilization of large pre-trained language models. These sophisticated models represent a significant leap forward, primarily due to their ability to understand and generate human-like text based on training from extensive datasets. Typically, these models are trained using unsupervised learning techniques, where they learn to predict the next word or token in a sequence by examining the tokens that precede it. This method has propelled them to the forefront of various NLP applications, including chatbots \citep{LLMchat}, text summarization \citep{LLMsum1,LLMsum2}, and advanced information extraction tasks \citep{LLMinex}.
%add references to the application of NLP in bioinformatics
Among the intriguing avenues explored with these models is their application in the field of bioinformatics, specifically in protein generation \citep{madani2020progen}. Indeed, the protein alphabet is composed of twenty common amino acids, each represented by a single character. Regarding their primary structure, proteins, which are vital biological molecules, are made up of chains of amino acids,  thus forming sequences of letters and drawing a parallel to the structure of natural languages. As in natural languages, protein sequences have directionality and are typically composed of reused modular elements that exhibit slight variations. Moreover, common protein motifs and domains, which are the basic building blocks of proteins, are similar to words, phrases, and sentences in human language.
This %sequence 
similarity suggests that language models, which excel in handling sequential data, could effectively generate amino acid chains, or proteins.\\
The primary objective of our research lies in advancing the understanding and application of medium-sized language models, particularly those in the 7 billion to 8 billion parameter range, including \texttt{Mistral-7B}, \texttt{Llama-2-7B}, \texttt{Llama-3-8B}, and \texttt{gemma-7B}, for the generation of high-quality protein sequences. Our hypothesis, backed by preliminary studies, suggests that these models, even when trained with considerably small datasets, can produce accurate and viable protein sequences effectively.\\
Furthermore, we extend our investigation to encompass a comparative analysis utilizing established protein-focused language models such as ProGen (\citet{nijkamp2023progen2, madani2020progen}), ProtGPT2 (\citet{ferruz2022protgpt2}), and ProLLaMA (\citet{lv2024prollama}). By applying the same experimental conditions across different models, we aim to provide quantitative and qualitative comparisons of their performance and effectiveness.\\
Ultimately, this study seeks to validate the capability of medium-sized models in protein design, emphasizing the potential of employing more compact, cost-efficient language models as powerful tools in bioinformatics research. This approach may significantly expedite scientific research and practical applications, spanning from drug design to precision medicine to other interdisciplinary fields.

This paper makes the following contributions:

\begin{itemize}
    \item \textbf{Exploration of Medium-sized LLMs --} We investigate the efficacy of medium-sized language models, with 7-8 billion parameters, in generating functionally viable protein sequences;
    \item \textbf{Adaptation to Small Data Sets --} We show that these models can achieve high performance even when trained with small datasets;
    \item \textbf{Comparative Analysis --} We provide a thorough comparative analysis of the performance of our models against established models in the field under identical experimental conditions;
    \item \textbf{Accessibility of Trained Models --} We commit to making all four trained language models developed for this study available to the scientific community to encourage further research and development.
\end{itemize}

The layout of this document is as follows: Section~\ref{sec:relatedworks} reviews previous research. Our methods are detailed in Section~\ref{sec:methodology}. Experimental results are discussed in Section~\ref{sec:experiments}, while conclusions and future perspectives are collected in Section~\ref{sec:conclusion}.
%%add some information about this work! and the objectives!!

\section{Related Works}\label{sec:relatedworks}
The integration of natural language processing (NLP) techniques, traditionally applied to human languages, into bioinformatics, has transformative potential, particularly in the analysis of biological sequences such as DNA, RNA, and proteins. These biological data, sharing similarities with linguistic texts in their structured and functional building blocks, are highly amenable to computational methodologies. The impactful success seen in %linguistic 
NLP through transformer-based models has led to breakthroughs in specialized models geared toward understanding the complexities of these biological sequences. By utilizing extensive databases such as UniProt \citep{uniprot2019uniprot}, ENSEMBL \citep{cunningham2022ensembl}, and GenBank \citep{benson2012genbank}, these models harness rich data to enhance both predictive and analytical capabilities in bioinformatics.\\
The realm of protein sequences has seen notable advancements through the adoption of both supervised and unsupervised learning models. Language models have been increasingly leveraged and employed in the domain of protein design \citep{ferruz2022controllable}. Supervised learning approaches refine models by training them with labeled data, which is invaluable for accurately predicting protein stability or identifying structural similarities among sequences \citep{bepler2021learning,alley2019unified}. On the other hand, the introduction of transformer technology has been pivotal in popularizing unsupervised learning methods \citep{vaswani2017attention}. These methods involve the strategic corruption of input sequences which are then used to train models to predict and reconstruct the NATURAL sequence. Leading models such as ESM \citep{rives2021biological}, ProtTrans \citep{elnaggar2021prottrans}, and ProteinBERT  \citep{brandes2022proteinbert} demonstrate this approach, offering powerful embeddings that prove critical in supporting a wide array of downstream biochemical tasks \citep{yang2024convolutions,rao2019evaluating}. These tasks include, but are not limited to, analyzing protein-protein interactions, predicting molecular functions, and identifying potential sites for drug binding. In addition to these developments, the adoption of autoregressive models — widely recognized for their ability to generate coherent, long-form text in classical NLP settings — has been successfully applied to the domain of protein sequencing. Prototypes like ProGen \citep{nijkamp2023progen2,madani2020progen}, ProtGPT2 \citep{ferruz2022protgpt2} and ProLLaMA \citep{lv2024prollama} capitalize on this capability, employing autoregressive algorithms to effectively predict the future elements of protein sequences from given contexts. This predictive ability is critical for sophisticated applications such as protein design, where the generation of novel and functionally effective proteins is required. \\
In this study, we employ some pre-trained language models, which we further fine-tune for protein generation tasks, by retraining both the tokenizers and the entire models. We then compare the results with those from other large language models (LLMs) currently available for protein generation tasks.
\section{Methodology}\label{sec:methodology}
\begin{figure*}[ht]
  \centering
  \includegraphics[width=\textwidth]{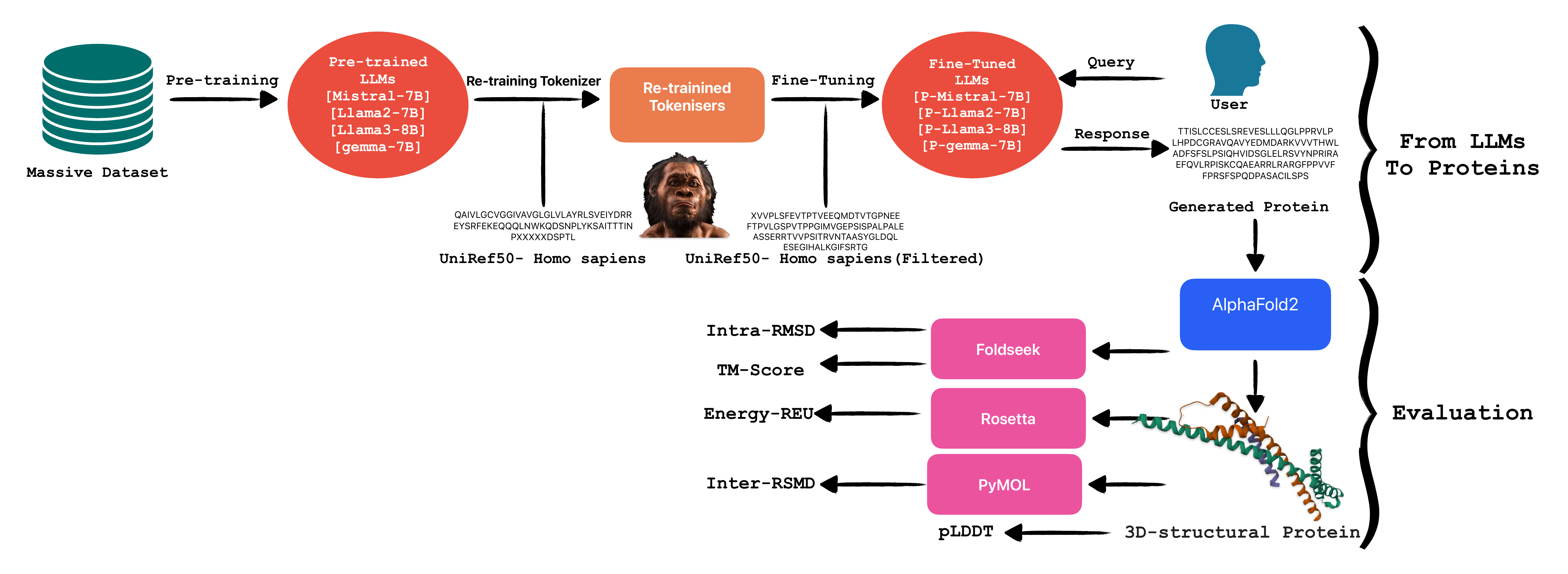}
  \caption{A comprehensive overview of our methodology employed for training, evaluating, and validating the protein sequence generation model. We initially retrained tokenizers for four distinct large language models --- \texttt{Mistral-7B}, \texttt{Llama-2-7B}, \texttt{Llama-3-8B}, and \texttt{gemma-7B} --- using the UniRef50-Homo sapiens dataset employing the Byte-Pair Encoding (BPE) technique. Subsequently, we fine-tuned these models on a filtered subset of the UniRef50-Homo sapiens dataset, aiming to minimize the loss associated with predicting subsequent protein sequences. For evaluation,  model output was validated using AlphaFold 2 to construct 3D protein structures, followed by assessments of the generated protein structural accuracy using metrics such as per-residue confidence score (pLDDT) from AlphaFold 2, RMSD (Root Mean Square Deviation), and TM-Score to compare topological similarities with known protein structures applied using FoldSeek. Additional evaluation included the use of Rosetta- Relax for analyzing the energetic profiles of the generated proteins. Finally, protein structural comparisons within each dataset were conducted using PyMOL to calculate the intra-dataset RMSD.}
  \label{fig:fig1}
\end{figure*}

In this section, we delineate the methodologies employed to adapt pre-trained LLMs for the generation of protein sequences. Our approach involved refining the tokenizer based on the Byte-Pair Encoding (BPE) methodology, followed by fine-tuning the entire pre-trained model using a designated dataset of protein sequences. Subsequently, this fine-tuned model was utilized to generate new protein sequences. 
It's important to note that base models such as LLMs, while powerful, are not inherently capable of designing novel proteins. Their success in this domain is achieved through a specialized fine-tuning process, which involves not only adapting the model to a specific task using a smaller, task-specific dataset but also modifying the tokenizer. This is because the tokens that LLMs were initially trained on are natural language tokens, whereas our domain requires a different set of tokens. Therefore, we also need to train the tokenizer to handle this new domain effectively.
Verification of these sequences was carried out by generating their respective PDB structures using DeepMind's AlphaFold  \cite{jumper2021highly}. We assessed the quality of these structures using various metrics such as pLDDT, RMSD, TM-Score and REU. The performance of the models --- namely \texttt{Mistral-7B} \cite{jiang2023mistral}, \texttt{Llama-2-7B} \cite{touvron2023llama}, \texttt{Llama-3-8B} , and \texttt{gemma-7B} \cite{team2024gemma} --- was then compared with previous studies that employed language models for protein sequence generation.
We have also evaluated the potential fitness of our generated sequences in comparison to natural and random sequences in the context of pLDDT, Rosetta-Relax scores, RMSD and TM-Scores, thereby providing a comparative analysis.
Figure \ref{fig:fig1} illustrates this methodology. Subsequently, we will provide a detailed description of all these steps, focusing on the training of the LLMs and their validation.

\subsection{From LLMs to Proteins}
Large language models, such as transformers, are sophisticated algorithms trained on extensive textual datasets. These models utilize their predictive capability primarily to determine the subsequent token based on the preceding ones. Given their training on a vast amount of text data, LLMs are highly adaptable and can be finely tuned for specialized tasks, including summarizing specific document types like legal texts. An interesting application of these models is in the domain of protein generation. Proteins, being amino acid sequences, differ significantly from standard text data. This difference necessitates the retraining of tokenizers to achieve more accurate tokenization for proteins, enhancing the model ability to recognize and predict relevant patterns in amino acid sequences. Following the retraining, these adapted tokenizers are used to refine the parameters of pre-trained LLMs. This fine-tuning process tailors the LLMs to predict protein sequences effectively by generating valid protein structures. In subsequent sections, we will elaborate on the methodologies applied for tokenizer retraining, describe the various LLMs utilized, and discuss their specific fine-tuning.
\paragraph{Tokenizer retraining}

 In situations where the corpus significantly diverges from that utilized during the initial training of a language model, it becomes imperative to retrain the model from scratch. This process necessitates adjusting the tokenizer to accommodate the nuances of the new dataset. A tokenizer serves the critical function of converting textual data into numerical representations suitable for computational processing by language models.\\
For the retraining of our tokenizer, we employed the Byte-Pair Encoding (BPE) method. BPE is a hybrid between a character-level and word-level tokenizer. It starts with a base vocabulary of individual characters and iteratively merges the most frequently adjacent pairs of characters or character sequences. Through this methodology, BPE effectively manages the vocabulary size, allowing for efficient handling of unknown words by breaking them down into recognizable subwords. This is particularly beneficial in managing morphologically rich languages or corpora with specialized jargon.\\
In our adaptation process, we retained the original vocabulary size of the tokenizer used in prior models to maintain consistency and optimize integration with the pre-trained configurations. This approach ensures that the retrained models sustain compatibility with existing frameworks while benefiting from a tokenizer that is fine-tuned to the specific features of the new dataset.

\paragraph{Fine-Tune Pre-trained LLMs}
In this research, our objective was to assess the capabilities of various pre-trained language models in the specialized task of protein generation. To this end, we fine-tuned four distinct models: \texttt{Mistral-7B}, \texttt{Llama-2-7B}, \texttt{Llama-3-8B}, and \texttt{gemma-7B}. Each model is based on the transformer architecture, which is renowned for its effectiveness in handling sequence-to-sequence tasks and operates under a causal framework conducive to generative tasks.\\
The four models were specifically chosen to represent a bandwidth of computational capacities predominantly ranging between 7 billion and 8 billion parameters, enabling a focused analysis on how parameter scale influences model performance in biological sequence generation. \texttt{Mistral-7B}, developed by MistralAI, contains precisely 7 billion parameters. In contrast, both \texttt{Llama-2-7B} and the newer \texttt{Llama-3-8B} are products from Meta, featuring 7 billion and 8 billion parameters, respectively. The latter represents an advanced iteration within the LLama series, potentially offering enhancements in learning efficiency and output refinement. Finally, \texttt{gemma-7B} from Google, also with 7 billion parameters, extends our model diversity, providing an additional perspective from another leading tech giant’s approach to language model development.\\
By employing these models, we aim to conduct a thorough comparative analysis, examining not just the quantitative outcomes in terms of accuracy and efficiency in protein generation, but also qualitative aspects such as the fidelity and usability of generated sequences. Given the similar parameter size, any observed differences in performance can be more directly attributed to architectural nuances and training methodologies between the models. This study not only advances our understanding of the capabilities of high-capacity language models in biosciences but also guides future developments in computational biology and the deployment of AI-driven tools for scientific discovery.\\
Firstly, we observe that each of these language models employs variants of the cross-entropy loss function. Throughout the fine-tuning process, the objective is to minimize this loss, which effectively maximizes the probability of predicting subsequent tokens accurately, based on the context provided by previous tokens. This optimization directly enhances the model ability to generate coherent and contextually appropriate text.\\
Given a sequence of tokens, the cross-entropy loss 
%for a language model 
predicts the probability of each subsequent token based on the previous context, i.e., given
%. This is computed over a sequence of tokens 
\( x_1, x_2, ..., x_n \) in training data, 
%where 
the model is able to predict each subsequent token \( x_{t+1} \) 
based on previous tokens \( x_1, ..., x_t \). The formula for the loss across an entire sequence of length \( N \) is:
\[
\mathcal{L} = -\sum_{t=1}^{N} \log(p_{\text{model}}(x_{t+1} \mid x_1, x_2, ..., x_t))
\]
where \( p_{\text{model}}(x_{t+1} \mid x_1, ..., x_t) \) is the probability assigned by the model to the correct next token \( x_{t+1} \), conditioned on the sequence 
%of previous tokens 
\( x_1, ..., x_t \).\\
This loss not only encourages the correct prediction of the next token but also indirectly learns the contextual dependencies among the tokens in the sequence, which is crucial for the generation of coherent and contextually appropriate outputs in language models.
\subsection{Evaluation}
\begin{figure}[ht]
  \centering
  \includegraphics[width=\columnwidth]{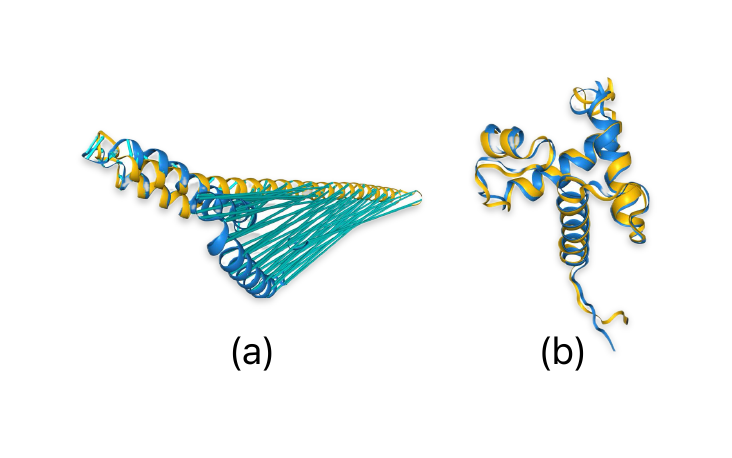}
  \caption{Illustration of High TM-Score and low Intra RMSD Compared to Low TM-Score and high Intra RMSD}
  \label{fig:tm-score}
\end{figure}

In this section, we describe each evaluation method implemented in our study following the generation of proteins. Initially, protein sequences generated using tuned LLMs were structurally modeled using   AlphaFold2\footnote{\href{  https://deepmind.google/technologies/alphafold/      }{deepmind.google/technologies/alphafold/ }}, which provided three-dimensional structures along with per-residue confidence scores (pLDDT). Subsequently, the topological similarity of these structures to known protein configurations was assessed using the TM-Score computed by FoldSeek \footnote{\href{ https://search.foldseek.com/  }{search.foldseek.com/ }}Additionally, Rosetta -Relax \footnote{\href{ https://www.rosettacommons.org/software }{www.rosettacommons.org/software }}was employed to analyze the energetic profiles of the modeled proteins, enhancing our understanding of their stability and viability. For intra-dataset structural comparisons, RMSD calculations were conducted using PyMOL\footnote{\href{ https://pymol.org/  }{pymol.org/ }} . Detailed descriptions and analyses of these metrics are provided in the following sections.
\paragraph{Alphafold2 (pLDDT)}
In the initial phase of the evaluation, we utilized AlphaFold2 to predict the structures of the generated proteins and compute their predicted Local Distance Difference Test (pLDDT) scores. AlphaFold2, developed by DeepMind, represents a significant advancement in protein structure prediction by leveraging sophisticated deep learning methodologies. It predicts protein structures from amino acid sequences, using extensive training datasets of known protein structures and incorporating a self-attention mechanism.  %This results in highly accurate predictions accompanied by pLDDT scores ranging from 0 to 100, which serve as per-residue confidence levels of the predicted structures. 
Moreover, pLDDT scores can be obtained, which provide valuable insight into structural accuracy, with values below 50 indicative of disordered regions, scores between 50 and 90 suggesting regions with some order, and scores above 90 denoting well-ordered regions.\\
%In this study, we developed and evaluated four novel protein generation models: Mistral, Gemma, Llama, and Llama3. These were compared against previously established models—ProLLama, ProtGPT, and ProGen—as well as original and random protein sequences.
\paragraph{Foldseek (TM-Score, Intra RMSD)}
To evaluate the accuracy of predicted protein structures, we utilized Foldseek, a robust tool designed for the comparison and analysis of three-dimensional protein structures. Foldseek is a tool for searching a set of query protein structures through a set of target protein structures. It uses a fast and sensitive k-mer and ungapped alignment prefilter from MMseqs2 on the 3Di sequences of the query and target structures to quickly identify candidate structures that are similar to the query. By submitting our predicted protein models to Foldseek, we computed two critical metrics: the TM-score and Root Mean Square Deviation (RMSD). The TM-score, ranging from 0 to 1, quantifies the global topological similarity between two protein structures, with higher scores indicating greater structural resemblance. Specifically, a TM-score above 0.5 generally indicates that the structures share the same fold, while a score below 0.3 suggests random structural similarity. Conversely, RMSD is a widely used metric in structural biology that assesses the similarity between two protein structures by comparing the positional differences of corresponding atoms, typically those in the backbone, after optimal superimposition. This metric provides insight into structural similarity from the perspective of atomic distances. In this study, we refer to this measure as 'Intra RMSD,' emphasizing the comparison between each predicted model and its respective known structure.A lower score is generally more desirable\\
Figure \ref{fig:tm-score} (a) illustrates an instance where the generated protein structure has limited similarity to the protein structure matched by Foldseek, as indicated by the green line in the figure. The protein structure in Figure \ref{fig:tm-score} (a) achieves a relatively low TM-Score of 0.28, indicating a weak resemblance to the matched protein structure. Furthermore, the substantial RMSD of 26.2 Å highlights a significant deviation and misalignment between the generated and matched structures.
In contrast, Figure \ref{fig:tm-score} (b) showcases a successful example of protein structure generation, where the generated structure has a high degree of similarity to the matched protein structure. The generated protein structure attains a high TM-Score of 0.90, signifying a strong structural similarity to the matched protein. Additionally, the low RMSD of 1.55 Å suggests that the generated structure has a high degree of precision and alignment with the matched structure.

\paragraph{Rosetta-Relax (REU)}
To comprehensively assess the quality of our predicted protein structures, we initiated the process by relaxing the native template. This initial relaxation ensures that the structure is energetically optimized from the outset, facilitating more accurate subsequent evaluations. Following the relaxation of the native template, we applied Rosetta-RelaxBB across all datasets. Rosetta-RelaxBB employs a Monte Carlo optimization approach that explores a range of backbone and rotamer conformations to minimize the Rosetta Energy function, which is based on biophysical principles and constraints. During each design iteration, amino acid side chains are substituted while maintaining fixed carbon backbone torsions. Energy minimization and relaxation are performed after threading the amino acid sequence through the known structure, allowing the backbone to transition into a potentially more stable energy state. Conformers with lower Rosetta Energy values indicate more relaxed and stable structures. The latest Rosetta Energy Forcefield (REF2015) shows a strong correlation with experimental parameters such as heat capacity, density, and enthalpy, providing a robust scoring function indicative of the thermodynamic stability of protein conformations.For a refined structure of this size, a score of -100 REU to -300 REU is typical. The lower the score, the more stable the structure is likely to be for a given protein. 

\paragraph{PyMOL (Inter RMSD)}
For the fourth phase of our evaluation, we utilized PyMOL, a sophisticated molecular visualization software equipped with extensive tools for protein structure analysis and comparison. PyMOL's features facilitate detailed examination of molecular structures and enable various quantitative assessments, such as calculating the Root Mean Square Deviation (RMSD). Specifically, we determined the Inter RMSD, which quantifies the RMSD for each trajectory within our datasets.As previously mentioned a lower score is generally more desirable.
\section{Experimental results}\label{sec:experiments}
In this section, we delineate the experiments conducted in this study, presenting an evaluation of the results garnered from the protein sequences we generated. Additionally, we discuss the regeneration of proteins utilizing language-based models specifically designed for protein generation tasks, including ProGen in four distinct sizes, ProtGPT2, and ProLLaMA.\\
Initially, we explore the dataset utilized in our experiments, which is notably smaller than those used in other models, followed by a detailed exposition of our training setup. Finally, we present a comprehensive analysis of the evaluation results employing various metrics such as pLDDT, RMSD, TM-Score and REU.
\subsection{Dataset}

In this study, the \textbf{UniRef50} dataset, originating from the UniProt databases, has been utilized. The UniProt Reference Cluster (UniRef) databases systematically organize clustered sets of protein sequences from UniProtKB \footnote{https://www.uniprot.org} and selected UniParc records, aiming to reduce redundancy and provide comprehensive coverage of sequence space. This is achieved through varying levels of sequence identity across three datasets, facilitating faster similarity searches among proteins.\\
Specific attention was given to the Homo sapiens subset within UniRef50, which initially comprised over 60,000 protein sequences. Given the constraints of computational resources and the criteria of our intended language models, a sequence length filter was applied. Only sequences below 512 tokens, as determined by our pre-trained tokenizer, were retained, narrowing the pool to 60,000  sequences.\\
For training and evaluation purposes, 42,000 sequences were allocated to the training set while the remaining 1,480 were designated for testing.\footnote{\href{https://huggingface.co/datasets/Kamyar-zeinalipour/UniRef50-HumanProteins}{huggingface.co/datasets/Kamyar-zeinalipour/UniRef50-HumanProteins/settings}}\\
This careful selection and allocation of sequences effectively optimized our computational resources and facilitated robust training and validation of our predictive models on protein sequences.
\begin{figure*}[ht]
  \centering
  \includegraphics[width=0.9\textwidth]{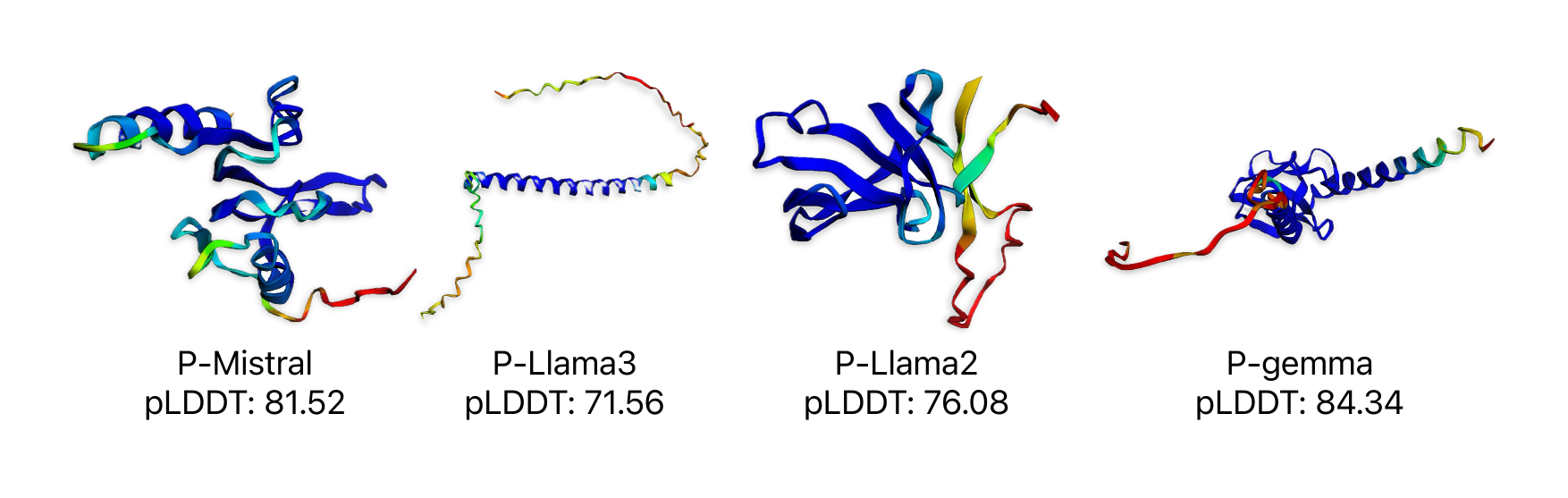}
  \caption{Examples of the 3D structure of proteins generated by each introduced model }
  \label{fig:example-genprot}
\end{figure*}

\begin{table*}[ht]
    \centering
    \begin{tabular}{llllllll}
        \hline
       Model & \#param & train &pLDDT $\uparrow$ & Intra$\downarrow$ & REU $\downarrow$ & TM $\uparrow$  & Inter $\downarrow$\\
       &  &size & & RMSD& & -Score2& RMSD \\
        \hline
      \texttt{NATURAL} & -- &   --& 67.77 & -- & -153.06 & -- & 4.40 \\
    \texttt{RANDOM}     & --&   --& 39.71 & 9.88 & -197.22 & 0.41 & 6.81 \\
   \texttt{P-Llama2}    & 7B&  42K&65.39 & 7.02 & -153.31 & 0.63 & 4.76 \\
   \texttt{P-Llama3}   & 8B&   42K& 62.99 & 7.38 & -132.50 & 0.65 & \textbf{4.30} \\
   \texttt{P-Mistral}    & 7B&  42K &\textbf{72.03}  & \textbf{5.42} & -197.40 & \textbf{0.68 }& 4.70 \\
    \texttt{P-gemma}     & 7B&   42K&62.24 & 5.80 & -141.60 & 0.65 & 5.83\\
    \texttt{PROLLAMA}      &7B &  -- & 55.80 & 9.46 & -126.65 & 0.47 & 5.66 \\
     \texttt{PROTGPT2}   & 774M&  49.8M & 64.50 & 6.52 & -146.23 & 0.52 & 5.52 \\
      
    \texttt{PROGENSMALL}   & 151M&  280M & 58.35 & 11.46 & -212.22 & 0.48 & 6.76\\
    \texttt{PROGENMEDIUM}    & 764M &   280M  & 58.98 & 11.64 & -240.89 & 0.59 & 11.20 \\
     \texttt{PROGENLARGE}    & 2.7B&   280M  & 61.78 & 7.65 & -158.18 & 0.58 & 5.47 \\
     \texttt{PROGENXLARGE}    & 6.4B&  280M   & 68.04 & 10.37 & \textbf{-251.37} & 0.54 & 6.05 \\

        \hline
    \end{tabular}
    \caption{Mean of the analyzed Metrics for each model.}
    \label{tab:mean_Metrics}
\end{table*}

\begin{figure*}[ht]
  \centering
  \includegraphics[width=0.9\textwidth]{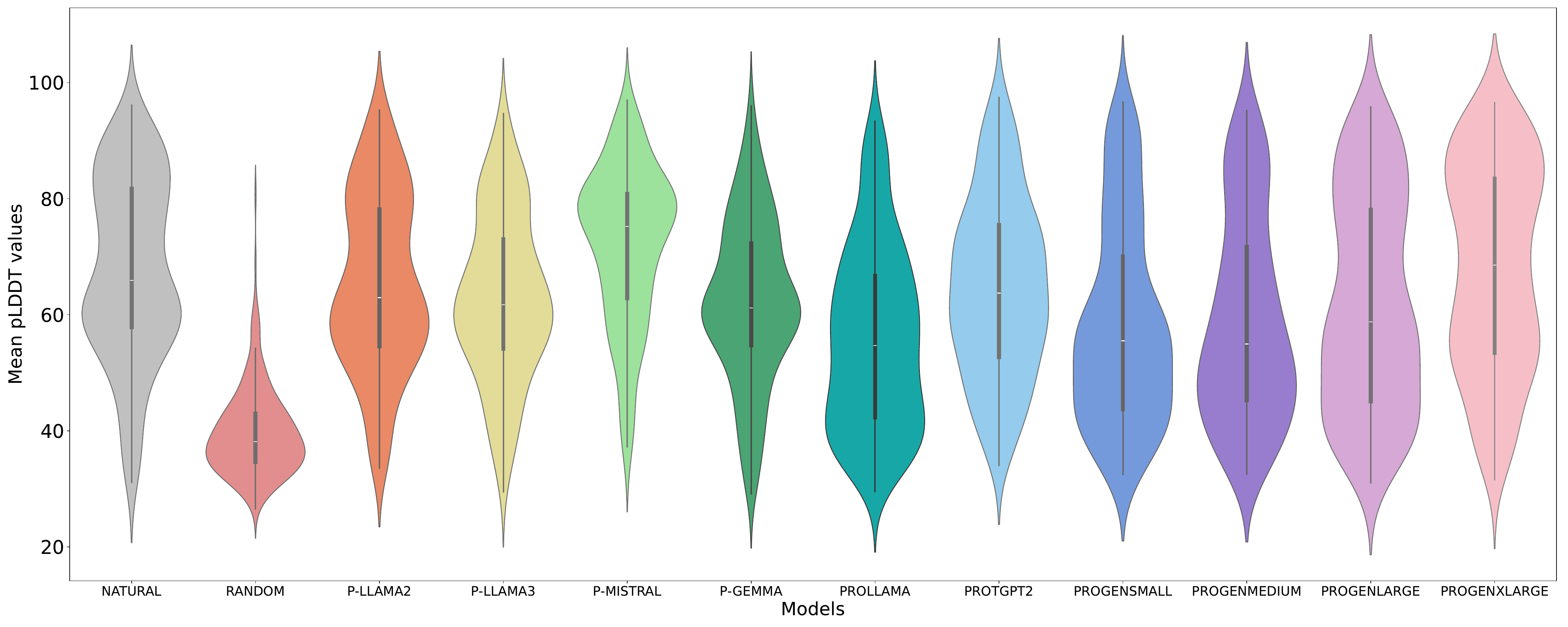}
  \caption{Violin plot of pLDDT   }
  \label{fig:fig2}
\end{figure*}

\subsection{Training Setup}

The training methodology employed in this study involved training Language Models (LMs) specifically tailored for protein generation utilizing four Nvidia A6000 GPUs. The training configuration utilized a sequence length of 512, with a maximum training step limit of 2000 and a batch size of 1, coupled with a gradient accumulation step size of 16 for enhanced training efficiency. The learning rate was set at 5e-5, and a cosine learning rate scheduler was employed to adaptively adjust the learning rate. Furthermore, a weight decay of 0.01 and a num warm-up step value of 150 were applied to stabilize the training process. The utilization of the bfloat16 data format contributed to faster computation due to reduced precision, enhancing overall training performance. We employed DeepSpeed \cite{rasley2020deepspeed}, a deep learning optimization library developed by Microsoft, to facilitate efficient training and optimization of the models. and also we applied FlashAttention 2 \cite{dao2023flashattention}.\\ 
Four distinct LLMs models, namely \texttt{Mistral-7B}, \texttt{Llama-2-7B}, \texttt{Llama-3-8B}, and \texttt{gemma-7B} were trained using this meticulously tuned training configuration. The selection of appropriate hyperparameters and the utilization of multiple GPUs facilitated efficient and timely training of these models. The strategic incorporation of the cosine learning rate scheduler and weight decay mechanism bolstered the models' convergence and performance during training, ultimately leading to the successful generation of protein sequences.\\
\subsection{Results Evaluation}
In this section, we randomly selected 250 proteins, each with a length between 70 to 140 amino acids, from each of the under-investigation models for structure prediction and subsequent evaluation. In order to initiate the protein generation process, we input a special token, known as the beginning-of-sequence (BOS) token. Once this token is fed into the model, it begins to generate protein sequences, leveraging the patterns and knowledge it has acquired during its training phase. These proteins were submitted to AlphaFold2, which generated 3D structural models with corresponding pLDDT scores for each protein. Examples of these 3D structures and corresponding pLDDT can be seen in Figure  \ref{fig:example-genprot}.
We proceeded to randomly select 20 3D structural proteins from each of the under-investigated models for a more in-depth analysis. The chosen proteins were then subjected to further evaluations, including the calculation of Intra RMSD, Inter RMSD, TM-Score, and REU with selected proteins. This multi-faceted approach to evaluation has allowed us to thoroughly assess the performance of our models and the quality of our 3D protein structure predictions.\\
To evaluate the pLDDT score for each protein, AlphaFold2 generates five 3D structural models with corresponding pLDDT scores. We then calculated the mean of the five pLDDT scores to obtain a representative pLDDT score for each protein. We present the evaluation results using all the metrics discussed in Section \ref{sec:experiments}. Table \ref{tab:mean_Metrics} summarizes the mean values of each evaluation metric for each model. Notably, P-Mistral consistently outperforms all other models across various metrics. Detailed information on these metrics, as well as corresponding plots and tables, are provided in the Appendix \ref{sec:appendix}.\\
The most significant difference between the trained models and randomly generated proteins We procedurally generated a set of proteins in a random manner, with each of these proteins being composed of a sequence of 20 amino acids, is observed in the pLDDT metric, as depicted in Figure \ref{fig:fig2}. Our models,\texttt{P-Llama2}  and  \texttt{P-Llama3}, exhibit a distribution similar to the NATURAL data. Additionally, we observed a significant disparity between randomly generated proteins and other models when evaluating the TM-score metric, as illustrated in Figure \ref{fig:fig4}. Other metrics, such as Inter and Intra RMSD, are shown in Figures \ref{fig:fig5} and \ref{fig:fig3}.\\
Furthermore, for the REU metric, we identified an optimal range between -100 and -300. The randomly generated proteins fall significantly outside this interval, whereas the models we introduced predominantly fall within the same range as the NATURAL data, as seen in Figure \ref{fig:fig5}.
The most intriguing finding of our study is that we were able to achieve and even surpass the performance of models trained on massive protein datasets, using a significantly smaller dataset. This was demonstrated across various evaluation metrics.

\section{Conclusion}\label{sec:conclusion}

In this study, we introduced four novel models designed to generate high-quality protein sequences by leveraging pre-trained language models. This research is motivated by the growing demand for efficient and accurate tools that can assist in understanding and engineering protein structures, which are pivotal in numerous biological and medical applications. Our approach involved a meticulous design and training phase, followed by rigorous testing and validation processes to assess the performance of each model.\\
To provide a thorough evaluation, we conducted comprehensive experiments comparing our models with a range of existing models that also utilize language models for protein sequence generation. 
Comparative analyses were performed, which were grounded on diverse criteria, including sequence quality, diversity, and fidelity to biological functions. These analyses also incorporated several structural assessment metrics such as pLDDT (predicted Local Distance Difference), TM-Score (to assess structural similarity), RMSD (Root Mean Square Deviation), and REU (Rosetta Energy Unit).
Our findings revealed that some of our proposed models, particularly \texttt{P-Mistral}, exhibited superior performance compared to existing models, even surpassing those trained on considerably larger datasets. This remarkable performance underscores the potential of our models to offer significant advancements in the field of protein sequence generation.\\
We are committed to the principles of open science and reproducibility. Consequently, we will make all four models publicly available to the research community. This accessibility will empower other researchers to utilize and build upon our work, fostering further advancements in the field of protein sequence generation.\\
Moreover, We aim to extend the capabilities of these models by incorporating instruction tuning to generate proteins with specific constraints. This will involve refining the models to adhere to certain criteria, such as ensuring the sequences have particular structural or functional properties. Such advancements could be pivotal in various applications, including drug design, synthetic biology, and understanding protein interactions at a deeper level.
    While our current implementation of LLMs for protein generation excels in unconditional generation, there is a need to explore and develop methods for generating conditional proteins. This would allow us to guide the generation process toward specific protein characteristics or functions, thereby enhancing the practical utility of our model.

\section*{Acknowledgments}

The funding for this paper was provided by the TAILOR project and the HumanE-AI-Net projects, both supported by the EU Horizon 2020 research and innovation program under GA No 952215 and No 952026, respectively.
\bibliography{custom}

\appendix

\section{Appendix}
\label{sec:appendix}
In this appendix, we have included violin plots and descriptive statistics for all the evaluation metrics utilized throughout this study. The violin plots offer a visual representation of the distribution and density of the data, enabling an in-depth comparison between different models or methods. Additionally, the descriptive statistics provide a comprehensive summary of the central tendency, dispersion, and shape of the distribution of each metric, including measures such as mean, median, standard deviation, and interquartile range. These tools together facilitate a thorough understanding of the performance and variability of the metrics used, thereby supporting a robust assessment of the study results.

\paragraph{pLDDT}

The violin plot of the mean pLLDDTs of each model is shown in Figure \ref{fig:fig2}, while its descriptive statistics are collected in Table \ref{tab:pLDDT}.

\begin{table*}[ht]
    \centering
    \begin{tabular}{lllllll}
        \hline
        Model & Q1 & Q3 & mean & median & min & max\\
        \hline
        \texttt{NATURAL} & 57.85 & 81.69 & 67.77 & 65.93 & 31.1 & 96.10 \\
     \texttt{RANDOM} & 34.62 & 42.95 & 39.71 & 38.12 & 26.52 & 80.74\\
       
        \texttt{P-Llama2} & 54.55 & 78.15 & 65.39 & 62.95 & 33.58 & 95.26 \\
        \texttt{P-Llama3} & 54.12 & 73 & 62.99 & 61.71 & 29.46 & 94.68\\
       \texttt {P-Mistral} & 62.82 & 80.83 & 72.03 & 75.22 & 35.06 & 96.98 \\
        \texttt{P-gemma} & 54.76 & 72.26 & 62.24 & 61.17 & 29.1 & 96.00\\
         \texttt{PROLLAMA} & 42.3 & 66.68 & 55.80 & 54.7 & 29.52 & 93.34\\
        \texttt{PROTGPT2} & 52.73 & 75.45 & 64.50 & 63.72 & 34.02 & 97.46 \\
         \texttt{PROGENSMALL} & 43.68 & 70.09 & 58.35 & 55.51 & 32.46 & 96.68\\
        \texttt{PROGENMEDIUM} & 45.27 & 71.71 & 58.98 & 54.96 & 32.56 & 95.20\\
       \texttt{PROGENLARGE} & 45.06 & 78.14 & 61.78 & 58.78 & 31.06 & 95.84\\

        \texttt{PROGENXLARGE} & 53.43 & 83.45 & 68.04 & 68.56 & 31.56 & 96.52\\

        \hline
    \end{tabular}
    \caption{Summary statistics for pLDDT}
    \label{tab:pLDDT}
\end{table*}

\paragraph{TM-Score}

\begin{figure*}[ht]
  \centering
  \includegraphics[width=0.9\textwidth]{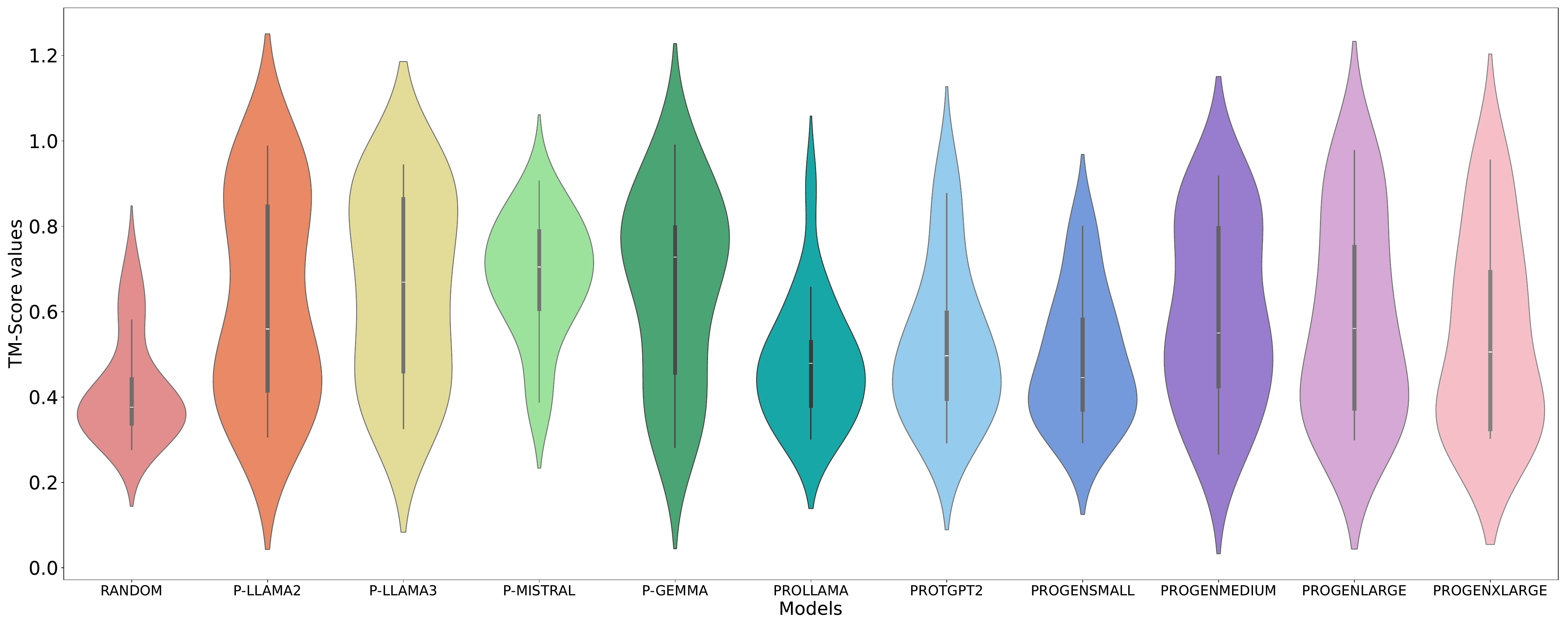}
  \caption{Violin plot of TM-Score  }
  \label{fig:fig4}
\end{figure*}

The violin plot of the TM-Score of each model is shown in Figure \ref{fig:fig4}, while its descriptive statistics are collected in Table \ref{tab:TM-Score}.

\begin{figure*}[ht]
  \centering
  \includegraphics[width=0.9\textwidth]{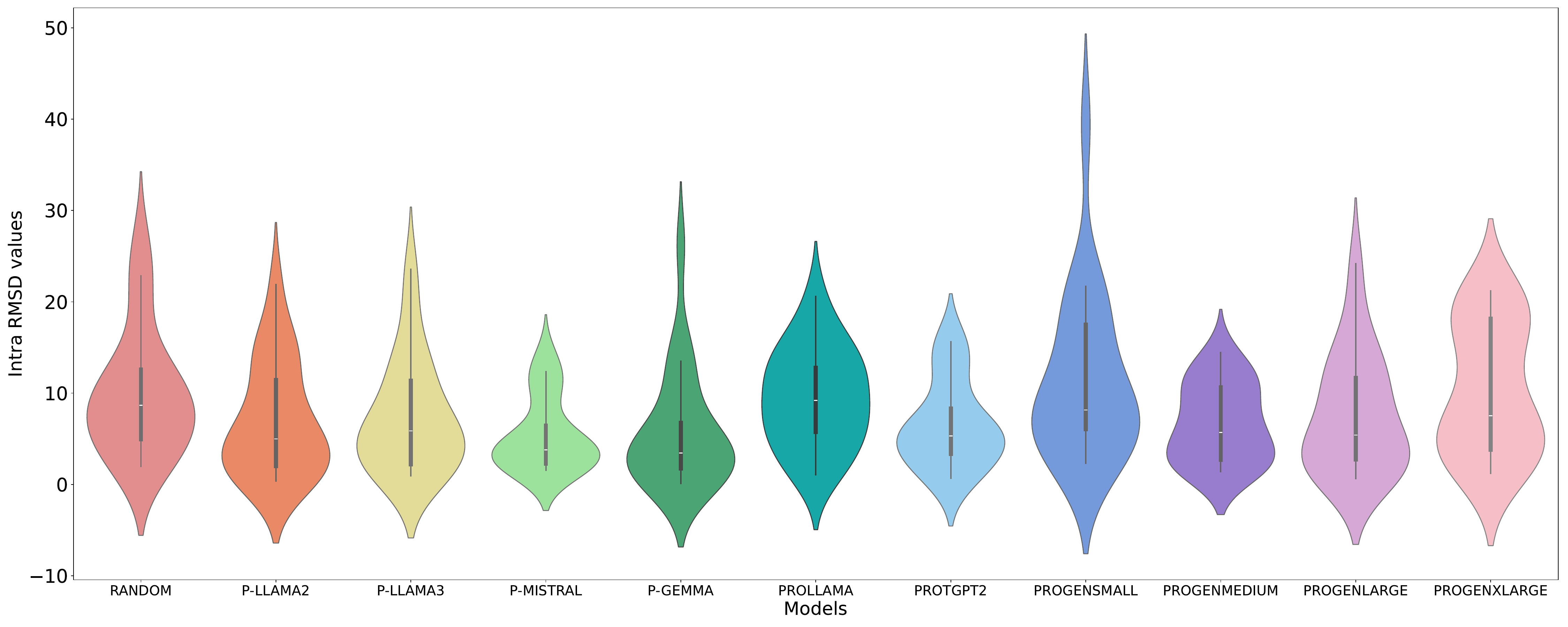}
  \caption{Violin plot of Intra RMSD   }
  \label{fig:fig3}
\end{figure*}

\begin{table*}[ht]
    \centering
    \begin{tabular}{lllllll}
        \hline
        Model & Q1 & Q3 & mean & median & min & max\\
        \hline
       \texttt{RANDOM} & 0.33 & 0.44 & 0.41 & 0.37 & 0.27 & 0.71 \\
       
        \texttt{P-Llama2} & 0.41 & 0.84 & 0.63 & 0.55 & 0.30 & 0.99 \\
        \texttt{P-Llama3} & 0.46 & 0.86 & 0.65 & 0.66 & 0.32 & 0.94\\
       \texttt{P-Mistral} & 0.60 & 0.78 & 0.68 & 0.70 & 0.38 & 0.91 \\
        \texttt{P-gemma} & 0.45 & 0.79 & 0.65 & 0.72 & 0.28 & 0.99\\
         \texttt{PROLLAMA} & 0.37 & 0.52 & 0.47 & 0.47 & 0.30 & 0.90\\
       \texttt{PROTGPT2} & 0.39 & 0.59 & 0.52 & 0.49 & 0.29 & 0.92\\
       \texttt{PROGENSMALL} & 0.37 & 0.58 & 0.48 & 0.44 & 0.29 & 0.80 \\
       \texttt{PROGENMEDIUM} & 0.42 & 0.79 & 0.59 & 0.55 & 0.26 & 0.92\\
       \texttt{PROGENLARGE} & 0.37 & 0.75 & 0.58 & 0.56 & 0.30 & 0.98 \\
        
        \texttt{PROGENXLARGE} & 0.32 & 0.69 & 0.54 & 0.50 & 0.30 & 0.96\\

        \hline
    \end{tabular}
        \caption{Summary statistics for TM-Score}
    \label{tab:TM-Score}
\end{table*}
\paragraph{Intra RMSD}
The violin plot of the Intra RMSD of each model is shown in Figure \ref{fig:fig3}, while its descriptive statistics are collected in Table \ref{table:intraRMSD}.

\begin{table*}[ht]
    \centering
    \begin{tabular}{lllllll}
        \hline
        Model & Q1 & Q3 & mean & median & min & max\\
        \hline
       \texttt{RANDOM }& 4.95 & 12.59 & 9.88 & 8.66 & 1.97 & 26.71\\
         \texttt{P-Llama2} & 2.01 & 11.43 & 7.02 & 4.98 & 0.39 & 21.88\\
       \texttt{P-Llama3} & 2.2 & 11.36 & 7.38 & 5.87 & 0.94 & 23.58 \\
       \texttt{P-Mistral} & 2.28 & 6.42 & 5.42 & 3.79 & 1.55 & 14.20\\
       \texttt{P-gemma} & 1.73 & 6.74 & 5.80 & 3.45 & 0.11 & 26.20\\
      \texttt{PROLLAMA} & 5.74 & 12.78 & 9.46 & 9.21 & 1.07 & 20.59\\
       \texttt{PROTGPT2} & 3.35 & 8.29 & 6.52 & 5.31 & 0.69 & 15.65 \\
        \texttt{PROGENSMALL} & 6.06 & 17.49 & 11.46 & 8.16 & 2.3 & 39.45\\
       \texttt{PROGENMEDIUM} & 2.68 & 10.62 & 11.64 & 5.69 & 1.42 & 14.46\\
       \texttt{PROGENLARGE} & 2.72 & 11.67 & 7.65 & 5.4 & 0.63 & 24.19\\

        \texttt{PROGENXLARGE} & 3.79 & 18.16 & 10.37 & 7.53 & 1.2 & 21.20\\

        \hline
    \end{tabular}
    \caption{Summary statistics for Intra RMSD}
    \label{table:intraRMSD}
\end{table*}

\begin{figure*}[ht]
  \centering
  \includegraphics[width=0.9\textwidth]{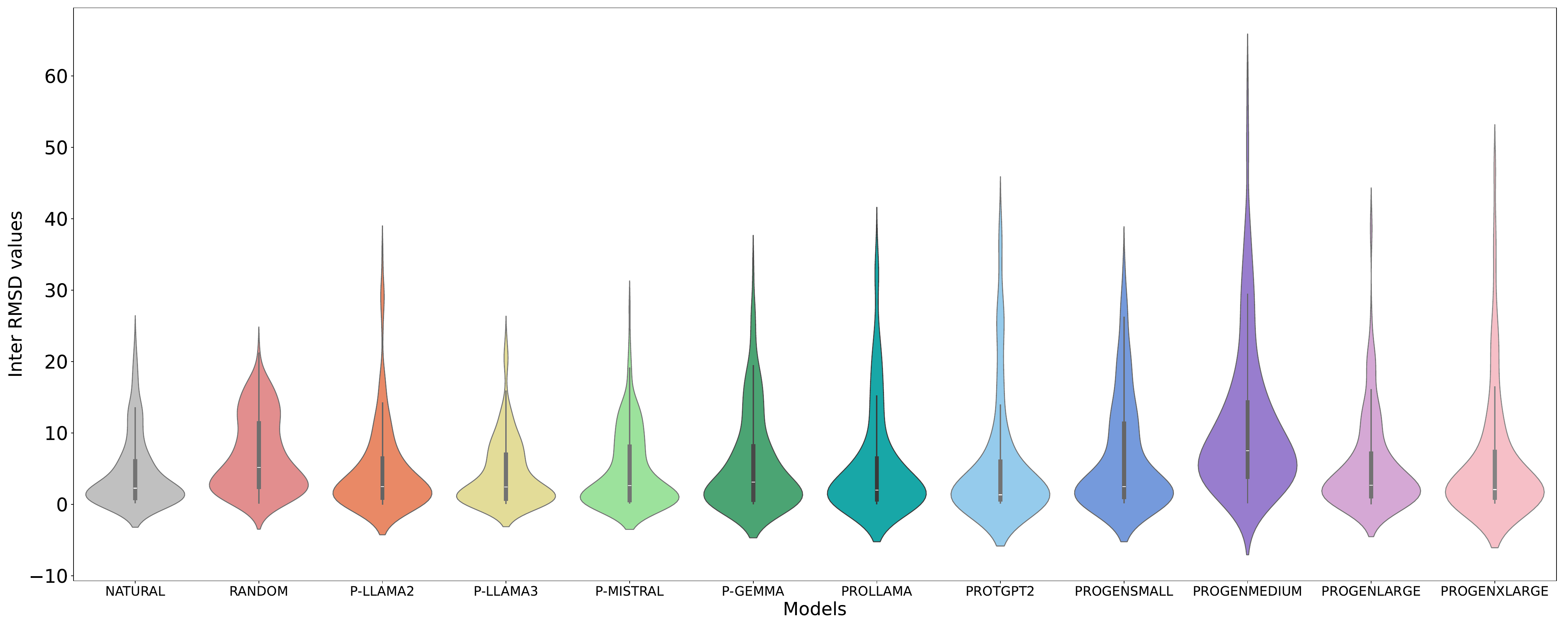}
  \caption{Violin plot of Inter RMSD }
  \label{fig:fig6}
\end{figure*}

\paragraph{Inter RMSD}
The violin plot of the Inter RMSD of each model is shown in Figure \ref{fig:fig6}, while its descriptive statistics are collected in Table \ref{tab:interRMSD}.

\begin{figure*}[ht]
  \centering
  \includegraphics[width=0.9\textwidth]{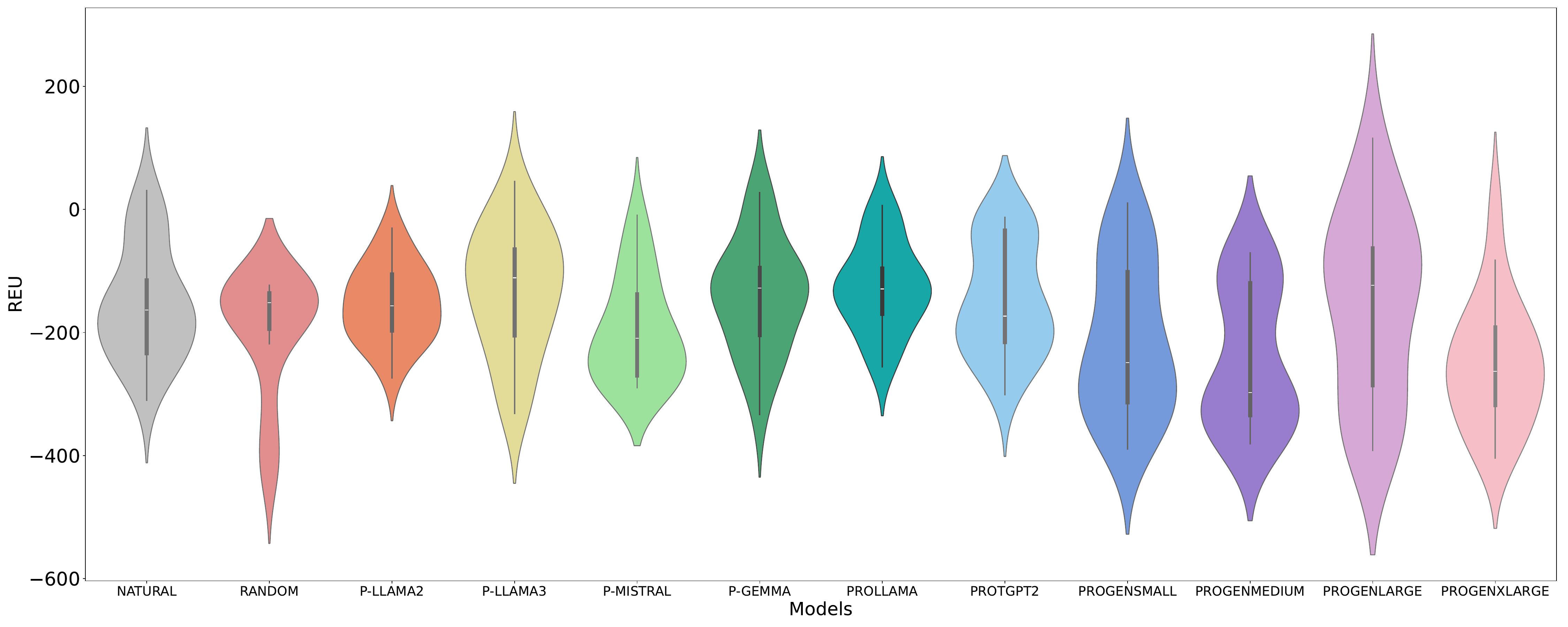}
  \caption{Violin plot of REU }
  \label{fig:fig5}
\end{figure*}

\begin{table*}[ht]
\centering
\begin{tabular}{lllllll}
\hline
Model & Q1 & Q3 & Mean & Median & Min & max \\
\hline
\texttt{NATURAL} & 0.87 & 6.04 & 4.40 & 2.24 & 0.23 & 22.98 \\
\texttt{RANDOM} & 2.41 & 11.34 & 6.81 & 5.15 & 0.16 & 21.18 \\
\texttt{P-Llama2} & 0.92 & 6.41 & 4.76 & 2.46 & 0.02 & 34.70\\
\texttt{P-Llama3} & 0.76 & 6.94 & 4.30 & 2.41 & 0.11 & 23.11 \\
\texttt{P-Mistral} & 0.55 & 8.08 & 4.70 & 2.61 & 0.18 & 27.59\\
\texttt{P-gemma} & 0.63 & 8.15 & 5.83 & 3.09 & 0.05 & 32.91  \\
\texttt{PROLLAMA} & 0.64 & 6.45 & 5.66 & 1.99 & 0.08 & 36.26 \\
\texttt{PROTGPT2} & 0.67 & 6.00 & 5.52 & 1.32 & 0.20 & 39.84 \\
\texttt{PROGENSMALL} & 1.02 & 11.29 & 6.76 & 2.45 & 0.20 & 33.43 \\
\texttt{PROGENMEDIUM} & 3.85 & 14.27 & 11.20 & 7.49 & 0.23 & 58.57 \\
\texttt{PROGENLARGE} & 1.10 & 7.11 & 5.47 & 2.66 & 0.07 & 39.71 \\

\texttt{PROGENXLARGE} & 0.88 & 7.37 & 6.05 & 2.06 & 0.16 & 46.93\\

\hline
\end{tabular}
\caption{Summary statistics for Inter RMSD}
\label{tab:interRMSD}
\end{table*}
\paragraph{REU}
The violin plot of the REU of each model is shown in Figure \ref{fig:fig5}, while its descriptive statistics are collected in Table \ref{tab:REU}.

\begin{table*}[ht]
    \centering
    \begin{tabular}{lllllll}
        \hline
        Model & Q1 & Q3 & mean & median & min & max\\
        \hline
        \texttt{NATURAL} & -233,48 & -115,43 & -153,06 & -163,30 & -310,23 &31.00 \\
        \texttt{RANDOM} & -193,85 & -136,43 & -197,22 & -151,57 & -434,25 & -123.39\\
        \texttt{P-Llama2} & -196,79 & -105,67 & -153,31 & -156,82 & -274,14 & -30.49\\
        \texttt{P-Llama3} & -205,03 & -65,05 & -132,50 & -111,15 & -331,44 & 45.39\\
       \texttt{P-Mistral} & -269,91 & -138,11 & -197,40 & -209,21 & -289,82 & -9.69\\
       \texttt{P-gemma} & -204,11 & -95,08 & -141,60 & -127,84 & -333,33 & 27.28\\
        \texttt{PROLLAMA} & -169,56 & -96,10 & -126,65 & -129,17 & -255,91 & 6.38\\
        \texttt{PROTGPT2} & -215,40 & -34,31 & -146,23 & -173,36 & -301,04 & -12.90\\
       \texttt{PROGENSMALL} & -313,85 & -101,59& -212,22 & -249,00 & -389,56 & 10.17 \\
        \texttt{PROGENMEDIUM} & -334,72 & -119,91 & -240,89 & -297,72 & -380,75 & -70.46\\
       \texttt{PROGENLARGE} & -286,13 & -63,01 & -158,18 & -123,29 & -391,96 & 115.96 \\

       \texttt{PROGENXLARGE} & -318,06 & -191,64 & -251,37 & -263,17 & -403,92 & 10.98\\

        \hline
    \end{tabular}
    \caption{Summary statistics for REU}
    \label{tab:REU}
\end{table*}

\end{document}